\documentclass[aps,prb,showpacs,reprint]{revtex4-1}
\usepackage{graphicx}
\usepackage{subfigure}
\usepackage{amsfonts}
\usepackage{amsmath}
\usepackage{amssymb}
\usepackage{subfigure}

\begin{document}

\title{Dynamics and decoherence in nonideal Thouless quantum motors}

\author{Lucas J. Fern\'andez-Alc\'azar$^1$}
\author{Horacio M. Pastawski$^1$}
\author{Ra\'ul A. Bustos-Mar\'un$^{1,2}$}
\thanks{Corresponding author: \texttt{rbustos@famaf.unc.edu.ar}}

\affiliation{$^1$Instituto de F\'{\i}sica Enrique Gaviola and Facultad de
Matem\'{a}tica, Astronom\'{\i}a, F\'{\i}sica y Computaci\'on, Universidad Nacional de
C\'{o}rdoba, Ciudad Universitaria, C\'{o}rdoba, 5000, Argentina}


\affiliation{$^2$Facultad de Ciencias Qu\'{\i}micas, Universidad Nacional de C\'{o}rdoba, Ciudad
Universitaria, C\'{o}rdoba, 5000, Argentina}


\begin{abstract}
Different proposals for adiabatic quantum motors (AQMs) driven by DC currents have recently attracted considerable interest.
However, the systems studied are often based on simplified models with highly ideal conditions where the environment is neglected.
Here, we investigate the performance (dynamics, efficiency, and output power) of a prototypical AQM, the Thouless motor.
To include the effect of the surroundings on this type of AQMs, we extended our previous theory of decoherence in current-induced forces (CIFs) to account for spatially distributed decoherent processes.
We provide analytical expressions that account for decoherence in CIFs, friction coefficients and the self-correlation functions of the CIFs.
We prove that the model is thermodynamically consistent and we find that decoherence drastically reduces the efficiency of the motor mainly due to the increase in conductance, while its effect on the output power is not much relevant.
The effect of decoherence on the current-induced friction depends on the length of the system, reducing the friction for small systems while increasing it for long ones.
Finally, we find that reflections of the electrons at the boundary of the system induce additional conservative forces that affect the dynamics of the motor.
In particular, this results in the hysteresis of the system and a voltage dependent switching.\end{abstract}

\maketitle

\section{Introduction}

In recent years, there has been an increasing interest in electronic transport through nanoelectromechanical systems (NEMS) where electronic degrees of freedom couple directly to the mechanical ones.\cite{Dundas,Bailey,nanocar,MolcecMotorsGral,NatureNano11,chiaravalloti2007rack}
Advances in the experimental control and fabrication of NEMS\cite{Craighead2000_NEMS,roukes2001_NEMS} as well in the theory describing current-induced forces (CIFs)\cite{vonOppenPRL11,vonOppenPRB12}  
has fueled a great variety of proposals in the field, such as nanocoolers, nanorefrigerators, and nanomotors among others.\cite{McEniry09current_cooling,arrachea2012refrigerator,galperin2009cooling,FastNM,AQM13}
An interesting example of these are the so called adiabatic quantum motors (AQMs).\cite{AQM13}
They are nanoscopic motors driven by the recoil forces of a flux of quantum particles.
There, quantum mechanics can be used  for example to boost their efficiency by enhancing electronic reflectance.
Essentially, this class of devices works as quantum pumps operated in reverse.
While in a quantum pump, the periodic movement of some parameters pumps quantum particles from one reservoir to another, in a quantum motor a DC current of particles induces the cyclic motion of the device.
The adjective ``adiabatic'' in this context refers to the limit when the dynamics of the mechanical degrees of freedom is slow compared with the dwell time of electrons passing through the device.
In this regime it is commonly assumed that the mechanical degrees of freedom behave classically.

An interesting example of AQMs is the one based on the Thouless pump.\cite{Thouless83,Zhang}
In this system, a mechanical degree of freedom couples to electrons through a periodic potential.
For large but finite conductors, a ``band gap'' arises reducing dramatically the transmission of electrons within an energy range.
As previously shown, this fact can be used in the so called Thouless motor to increase the efficiency of the engine.\cite{AQM13}
Since quantum interferences play a crucial role in the high performance of such systems, it is natural to wonder about the role of decoherent effects.\cite{FBP15,CFBP14} 
However, in this system, decoherent events can occur anywhere along the long conductor, and this should be properly taken into account.
Besides that, in the ideal Thouless motor, based on a linearized Hamiltonian which also neglects the effect of the boundaries of the system, conservative forces are negligible.
It is then relevant to study the deviations from the linearized model, the appearance of conservative forces, and their interplay with nonconservatives ones in the dynamics of the system.

In the present work, we use a tight-binding model to investigate the performance of AQMs based on the Thouless pump. We also include in our analysis the role of decoherence and the effect of the mismatching between the system and the leads.
In section \ref{sec_Theory} we present the theoretical framework used to describe the dynamics of mechanical systems subject to current-induced forces in presence of decoherence.
In subsection \ref{subsec_CIFs} we extend our previous theory of CIFs in presence of decoherence to account for spatially distributed decoherent processes.
In subsection \ref{subsec_System} we specify the models used to describe the Thouless motor and define its efficiency and output power.
The main results are discussed in section \ref{sec_Results}.
In subsection \ref{subsec_Deviations} we study the deviations of Thouless motors with respect to their 
ideal description.
In subsection \ref{subsec_Dynamics} we discuss the interplay of conservative and nonconservative forces in the dynamics of the system.
Finally, in subsection \ref{subsec_Decoherence} we investigate the effect of decoherence on the performance of these AQMs.

\section{Theory\label{sec_Theory}}
\textit{Langevin Dynamics - }
In this work we will treat the mechanical degrees of freedom of the system as classical fields acting on the electrons, treated as quantum spinless noninteracting particles. The dynamics of the classical fields are assumed to be slow compared to that of the electrons.
In this context of mechanical modes embedded in an electronic environment, it is natural to describe the dynamics of the classical system by a set of coupled Langevin equations,
\begin{equation}
M_\nu \ddot{x}_\nu +\frac{\partial U}{\partial x_\nu}=F_\nu - \sum_{\nu'} \gamma_{\nu,\nu'} \dot{x}_{\nu'} +\xi_\nu ,\label{eq_Langevin}
\end{equation}
where in the left hand side $M_\nu$ is the mass associated with coordinate $x_\nu$, and $dU/dx_\nu$ accounts for any force not included in the right hand side of the equation, which could be external forces or even internal forces but not classifiable as CIFs.
The right hand side of Eq. \ref{eq_Langevin} accounts for the CIFs\cite{vonOppenPRL11} where $F$ is
the mean adiabatic reaction force, or the Born-Oppenheimer force, and it is the result of the mean value of the force operator, $F_\nu = \left < \partial_\nu \hat H_{el} \right >$, for the frozen electronic Hamiltonian $\hat H_{el}$. 
The coefficients $\gamma_{\nu,\nu'}$ are associated with dissipative friction forces, for $\nu=\nu'$, and effective ``Lorentz'' forces, for $\nu\neq\nu'$, while $\xi_\nu$ accounts for the force fluctuations.
The dissipative term of the CIFs arises from the first adiabatic correction to the Born-Oppenheimer force, while fluctuations are given by the thermal and nonequilibrium contributions of the quantum fluctuations of the force.
All these terms have a quantum origin and expressions in terms of scattering matrices or Green's functions are given in Refs. \cite{vonOppenPRL11,Bode12}. The mean adiabatic reaction force can be written as
\begin{equation}
F_{\nu }=\sum_{\alpha}\int \frac{d\varepsilon }{2\pi \mathrm{i}} f_{\alpha}
\left( S^{\dagger } \frac{\partial S}{\partial x_{\nu }} \right)_{\alpha,\alpha} ,  \label{eq_Force1}
\end{equation}
where $S$ is the scattering matrix and $f_\alpha$ is the Fermi function of the conduction mode or channel $\alpha$ associated with some lead with chemical potential $\mu_\alpha$.
We will be interested only in the first order expansion of CIFs close to equilibrium. Then, it is enough to consider only the equilibrium contribution to $\gamma$, which is given by\cite{Bode12}
\begin{equation}
\gamma^{eq}_{\nu,\nu'} =
\frac{1}{2} \int^{}_{} \left. 
\frac{ d \varepsilon }{2\pi } \frac{\partial f}{\partial \varepsilon }
\sum_{\alpha,\beta}
\left(S^{\dagger }\frac{\partial S}{\partial x_\nu} \right)_{\alpha,\beta}
\left(S^{\dagger }\frac{\partial S}{\partial x_{\nu'}} \right)_{\beta,\alpha}
\right .
\label{eq_gamma_simet_frict}
\end{equation}
where $f$ is the equilibrium Fermi function.
Provided that electronic fluctuations occur at short time scales, $\xi_\nu$ is locally correlated in time and its variance $D_{\nu,\nu'}$ can be written as $\left\langle \xi_\nu(t)\xi_{\nu'} (t^{\prime })\right\rangle \approx D_{\nu,\nu'} \delta (t-t^{\prime })$, where its equilibrium value is given by
\begin{eqnarray}
D_{\nu,\nu'} = 2 k_B T  \gamma^{eq}_{\nu,\nu'}. \label{eq_DF}
\end{eqnarray}

\textit{Current -}
Suppose a system is connected to an arbitrary number of conduction channels $\alpha$ with chemical potential $\mu _{\alpha}$ and subject to the movement of an arbitrary number of classical degrees of freedom $x_\nu$ coupled somehow to the electronic degrees of freedom. Then, the total current at channel $\alpha$, in the adiabatic limit with respect to the movement of the ($x_\nu$)s, in the low bias regime, and for noninteracting particles, can be evaluated by
\begin{equation}
I_{\alpha}=I_{\alpha}^{bias}+I_{\alpha}^{pump}+\delta I_{\alpha},  \label{eq_current}
\end{equation}
where $I_{\alpha}^{bias}$ is the bias current consequence of the chemical potential differences among the channels, 
$I_{\alpha}^{pump}$ is the pumped current due to the adiabatic movement of the classical degrees of freedom, and $\delta I_{\alpha}$ is the noise in the current.

The bias current at channel $\alpha$ is given by\cite{Bt86PRB,PM01}
\begin{equation}
I_{\alpha}^{bias}=-\frac{e}{h}\sum_\beta T_{\alpha\beta}\delta \mu _{\beta},  \label{eq_I_bias}
\end{equation}%
where $e$ is the electron charge, $h$ the Planck constant, $T_{\alpha\beta}$ is the probability of transmission between channels $\alpha$ and $\beta$, i.e. an adimensional conductance \cite{IL99}, $T_{\alpha\alpha}=-\sum_{\beta \neq \alpha} T_{\alpha\beta}$, and $\delta \mu _{\beta}=\mu _{\beta}-\mu_0$, with $\mu_0$ being the equilibrium chemical potential. Note that we are not including the factor 2 in the current as we are considering that electrons with different spins belongs to different conduction channels of the same lead.

The pumped current at channel $\alpha$ can be evaluated by\cite{ButtPrTh94,Brower98}
\begin{equation}
I_{\alpha}^{~pump}=e\sum\nolimits_{\nu }\frac{\mathrm{d}n_{\alpha}}{\mathrm{d}x_{\nu }}%
\dot{x}_{\nu }, \label{eq_I_pump}
\end{equation}%
where $ dn_{\alpha}/dx_{\nu }$ is the emittance of channel $\alpha$ due to a change in the parameter $x_\nu$.\cite{book_capButtiker}
At equilibrium, it can be evaluated from
\begin{equation} 
\frac{\mathrm{d}n_{\alpha}}{\mathrm{d}x_{\nu }}=\int \frac{ d \varepsilon 
}{2\pi \mathrm{i}}\left( -\frac{\partial f}{\partial \varepsilon }%
\right)
\left(\frac{\partial S}{\partial x_{\nu }}S^{\dagger}\right)_{\alpha,\alpha}.
  \label{eq_emmissivity}
\end{equation}

Deviations from the mean current given by $\delta I_{\alpha}$ are provided by thermal (Johnson-Nyquist) noise and shot noise.\cite{ButtPhysRep} It is assumed that they consist of rapid fluctuations of the current where their current-current correlations are given by $\left\langle \delta \vec{I}_{\alpha}(t-t^{\prime })\delta \vec{I}_{\beta}(t^{\prime })\right\rangle=D_{\alpha,\beta}^{(I)}~\delta (t-t')$. The shot noise contribution to $\delta I_{\alpha}$ vanishes at equilibrium but thermal noise remains finite, resulting in \cite{ButtPhysRep}
\begin{equation}
D_{\alpha,\beta}^{(I)}=-2k_{B}T\frac{e^{2}}{h}T_{\alpha\beta}, \label{eq_DI}
\end{equation}%
where $k_{B}T$ is the thermal energy.

\textit{Decoherence - }
The type of decoherence modeled in this work considers that at any time interval ${\rm d}t$ electrons are locally removed from a site $i$ and with probability $2 \Gamma_\phi {\rm d} t/ \hbar$ from the coherent beam propagating through the system.
Each lost electron is then instantaneously reinjected with a random phase at the same position.\cite{GLBE1,GLBE2,Zimbovskaya,LucasPRA15}
In principle, the rate of decoherent processes, $1/\tau_\phi=2\Gamma_{\phi,i}/\hbar$, can be
estimated from the system-environment interaction through the Fermi
golden rule (FGR).
The interaction of electrons with phonons or other processes not included
in the original Hamiltonian, not only renormalizes the electron's
energy but also produces their decay towards an unbounded region of
Hilbert's space.
This results in an imaginary self-energy correction $\Gamma_{\phi,i}$
which we use as a free parameter.

As first noticed by B\"uttiker,\cite{Bt86PRB} a voltmeter connected to site $i$, being a classical apparatus, imposes the collapse of the wave function and results in a source of decoherence.
In the low bias limit, such a voltmeter is susceptible to be modeled by a multichannel scattering matrix supplemented with the condition of particle conservation imposed by the Kirchhoff's laws  that define the measured
voltage, i.e. $I_{\phi,i}=0$.
D'Amato and Pastawski\cite{DP90} noticed that local decoherent processes described by the FGR , and hence by a Hamiltonian description, can be modeled as leads attached to
specific sites.
In this case, the escape rate toward those fictitious leads is precisely the rate at which decoherent events occur, $2\Gamma_{\phi,i}/\hbar$, and this must be supplemented by the
condition $I_{\phi,i}=0$.
Current conservation is ensured by a local chemical potential at the fictitious voltage probes calculated self-consistently from the multi-terminal Landauer-B\"uttiker equations.
The advantage of this approach is that the values of ($\Gamma_{\phi,i}$)s can be estimated from the specific Hamiltonian model that describes the system-environment problem.\cite{CBMP10}

Later on, Pastawski proved that the fictitious probes approach results from the linear response approximation of the Keldysh-Kadanoff-Baym integro-differential formulation of quantum transport.
\cite{GLBE1,GLBE2}
In this local description, the energy of electrons is preserved but not its momentum which leads, in the limit of high decoherence rate, to a random-walk-like dynamics for long conductors.
This, in turn, provides the correct asymptotic limit for large conductors compatible with Ohm's law, $I \propto L^{-1}$ where $L$ is the length of the system \cite{DP90,GLBE1,GLBE2}.
Note that this can be used to estimate the value of $\Gamma_{\phi,i}$ from experiments.\cite{CBMP10} 
Direct comparison with dynamical simulations also showed that this steady state model of decoherent transport can also be seen as the effect of fast random fluctuations of local energies.\cite{LucasPRA15}

In a previous work, we extended the above mentioned method to account for decoherence in CIFs.\cite{FBP15} The key in this case is to consider all the terms of the current in Eq. \ref{eq_current}.
This result is consistent with the positivity of friction coefficients in equilibrium, Onsager reciprocity relations, as well as the fluctuation-dissipation theorem.
Such consistency is fundamental to prevent unphysical results.
In the following, we will further extend the method originally formulated for a two channel system of zero dimension, to include decoherence in a wider class of systems, those connected to an arbitrarily large number of reservoirs and subject to spatially distributed decoherent processes.

\subsection{Currents and current-induced forces for spatially distributed decoherent processes\label{subsec_CIFs}}

\textit{Currents -} In what follows, we will make a distinction between the real conduction channels and fictitious probes labeling them as $\ell$ and $\phi$, respectively. In this notation, the vector containing the total current at each channel,  $\vec{I}=\left( \vec{I}_{\ell }~,~\vec{I}_{\phi }\right)^T $, results in 
\begin{eqnarray}
\left(\begin{array}{c}
\vec{I}_{\ell } \\ 
\mathbf{0}%
\end{array}\right)
&=&-\frac{e}{h}%
\left(\begin{array}{cc}
\mathbb{T}_{\ell \ell } & \mathbb{T}_{\ell \phi } \\ 
\mathbb{T}_{\phi \ell } & \mathbb{T}_{\phi \phi }%
\end{array}\right)
\left(\begin{array}{c}
\delta \vec{\mu}_{\ell } \\ 
\delta \vec{\mu}_{\phi }%
\end{array}\right)
\nonumber \\ & &
+e\sum\nolimits_{\nu }%
\left(\begin{array}{c}
\mathrm{d}\vec{n}_{\ell }/\mathrm{d}x_{\nu } \\ 
\mathrm{d}\vec{n}_{\phi }/\mathrm{d}x_{\nu }%
\end{array}\right)
\dot{x}_{\nu }+%
\left(\begin{array}{c}
\delta \vec{I}_{\ell } \\ 
\delta \vec{I}_{\phi }%
\end{array} \right)
.  \label{eq_current_block}
\end{eqnarray}%
where we have used Eqs. from \ref{eq_current} to \ref{eq_emmissivity} and have included the null current condition through fictitious leads $\vec{I}_{\phi }\equiv 0$.\cite{CFBP14,beenakker1992suppression}
Explicit formulas for the calculation of $\mathbb{T}$ are given in references \cite{CFBP14} and \cite{CBMP10} among others.
By solving Eq. \ref{eq_current_block} by blocks, we obtain the $\delta \vec{\mu}_{\phi }$ that satisfy charge conservation,
\begin{eqnarray}
\delta \vec{\mu}_{\phi } &=& \delta \vec{\mu}_{\phi }^{\ neq}+\delta \vec{\mu}_{\phi }^{\ pump}+\delta \vec{\mu}_{\phi }^{ \ fluc}
\nonumber \\ &=&
\left( - \mathbb{T}_{\phi \phi }\right) ^{-1}
\mathbb{T}_{\phi \ell }~\delta \vec{\mu}_{\ell }
\nonumber \\ & &
- \sum\nolimits_{\nu } h\left( -\mathbb{T}_{\phi \phi }\right) ^{-1} \frac{\mathrm{d}\vec{n}_{\phi }}{\mathrm{d}x_{\nu }}\dot{x}_{\nu }
\nonumber \\ & &
- \frac{h}{e}\left( -\mathbb{T}_{\phi \phi }\right) ^{-1} \delta \vec{I}_{\phi } .  \label{eq_chem_pot_ML}
\end{eqnarray}
We consider that the last term $\delta \vec{\mu}_{\phi }^{ \ fluc}$ fluctuates with
a fast time scale, compared with that of the mechanical degrees of freedom.
Replacing $\delta \vec{\mu}_{\phi }$ into Eq. \ref{eq_current_block}, we obtain the expression for the electronic currents corrected by decoherent processes,
\begin{eqnarray}
\vec{I}_{\ell } &=&-\frac{e}{h}\left\{ \mathbb{T}_{\ell \ell }+\mathbb{T}%
_{\ell \phi }\left( -\mathbb{T}_{\phi \phi }\right) ^{-1}\mathbb{T}_{\phi
\ell }\right\} \delta \vec{\mu}_{\ell }
\nonumber \\
&&+e\sum\nolimits_{\nu }\left\{ \frac{\mathrm{d}\vec{n}_{\ell }}{\mathrm{d}%
x_{\nu }}+\mathbb{T}_{\ell \phi }\left( - \mathbb{T}_{\phi \phi }\right) ^{-1}%
\frac{\mathrm{d}\vec{n}_{\phi }}{\mathrm{d}x_{\nu }}\right\} \dot{x}_{\nu }
\nonumber \\
&&+\delta \vec{I}_{\ell }+\mathbb{T}_{\ell \phi }\left( -\mathbb{T}_{\phi
\phi }\right) ^{-1}\delta \vec{I}_{\phi }.  \label{eq_currents_complete}
\end{eqnarray}%
The first term is the ``bias'' current that includes decoherent corrections, also derived in Ref. \cite{CFBP14}.
The second one is the pumped current in the presence of decoherence which results in an extension of the expression derived in Refs. \cite{MkBt01,CrBrow02}.
Finally, the third term of Eq. \ref{eq_currents_complete} accounts for the currents fluctuations due to thermal noise and can be evaluated using Eq. \ref{eq_DI}.

\textit{Current induced forces - }
We start by splitting the Fermi function as $f_{\alpha}=f+\Delta f_{\alpha}$. 
Then, the force can be expressed as
\begin{equation}
F_{\nu }=F_{\nu }^{eq}+\Delta F_{\nu }^{{}}.
\end{equation}%
The equilibrium force $F_{\nu }^{eq}$, which is conservative, is given by
\begin{equation}
\vec{F}^{eq}=-\nabla U^{eq},
\end{equation}%
where
\begin{equation}
U^{eq}=-\int \frac{\mathrm{d}\varepsilon }{2\pi \mathrm{i}}%
f \mathrm{Tr}\left( \ln S \right). \label{eq_Ueq}
\end{equation}%
The other term of the force, $\Delta F_{\nu }$, is
\begin{equation}
\Delta F_{\nu }=\sum_{\alpha}\int \frac{\mathrm{d} \varepsilon }{2\pi \mathrm{i}}%
\Delta f_{\alpha}
\left(S^{\dagger }\frac{\partial S}{\partial x_{\nu }} \right)_{\alpha,\alpha}
.  \label{eq_Delta_F}
\end{equation}%
Taking the limit of low temperatures and small bias, one can rewrite Eq.
\ref{eq_Delta_F} as
\begin{equation}
\Delta F_{\nu }=\left ( \frac{\mathrm{d}\vec{n}^{(*)}}{\mathrm{d}x_{\nu }} \right )^T \cdot \delta \vec{\mu}
\label{eq_Delta_F_short}
\end{equation}%
where $\mathrm{d}\vec{n}^{(*)}/\mathrm{d}x_\nu =(\mathrm{d}\vec{n}^{(*)}_{\ell }/\mathrm{d} x_\nu ~,~\mathrm{d}\vec{n}^{(*)}_{\phi }/\mathrm{d} x_\nu )^T$, $\delta \vec{\mu}=(\delta \vec{\mu}_{\ell },\delta \vec{\mu}_{\phi })^T$, and
the injectance $(\mathrm{d}\vec{n}^{(*)} / \mathrm{d} x_{\nu })$, is given by\cite{book_capButtiker}
\begin{eqnarray}
\frac{\mathrm{d}n^{(*)}_{\alpha}}{\mathrm{d}x_{\nu }}(\vec{B})
&=&
\int \frac{ \mathrm{d}\varepsilon 
}{2\pi \mathrm{i}}\left( -\frac{\partial f}{\partial \varepsilon }%
\right)
\left(S^{\dagger}(\vec{B})\frac{\partial S(\vec{B})} {\partial x_{\nu}}\right)_{\alpha,\alpha}  
\nonumber \\ &=&
\frac{\mathrm{d}n_{\alpha}}{\mathrm{d}x_{\nu }}(-\vec{B}).
\end{eqnarray}
where $\vec{B}$ represents some external magnetic field. In the last step we have used the reciprocity relation $S_{\alpha,\beta}(\vec{B})=S_{\beta,\alpha}(-\vec{B})$, which is a consequence of particle conservation, $S^{\dagger}S=\mathbb{I}$, and time reversal for spinless particles $S(-\vec{B})S^*(\vec{B})=\mathbb{I}$.\cite{Butt88reciprocity}

By replacing $\delta \vec{\mu}_{\phi }$ from Eq. \ref{eq_chem_pot_ML} into Eq. %
\ref{eq_Delta_F_short}, we obtain 

\begin{eqnarray}
\Delta F_{\nu }&=&F_{\nu }^{ne}-\sum\nolimits_{\nu ^{\prime }}\gamma _{\nu
\nu ^{\prime }}^{\phi}\dot{x}_{\nu ^{\prime }}+\xi _{\phi,\nu }.
\label{eq_CIF_dec}
\end{eqnarray}
where the first term represents the nonequilibrium forces, the second gives the dissipative forces induced by decoherence, and the third provides the force fluctuations induced by decoherence. These last two terms add to the dissipative forces and force fluctuations already present in the system, Eqs. \ref{eq_gamma_simet_frict} and \ref{eq_DF}. 

The nonequilibrium forces are 
\begin{equation}
F_{\nu }^{ne}=\left\{ \frac{\mathrm{d}\vec{n}^{(*)}_{\ell }}{\mathrm{d}x_{\nu }}
+ \left ( \frac{\mathrm{d}\vec{n}^{(*)}_{\phi }}{\mathrm{d}x_{\nu }} \right )^T
\left( - \mathbb{T}_{\phi \phi }\right) ^{-1}\mathbb{T}_{\phi \ell }\right\}
\delta \vec{\mu}_{\ell }.  \label{eq_NEForces}
\end{equation}
These forces are not necessarily conservative, and its line integral over one cycle is proportional to the number of particles pumped over each channel,
\begin{equation}
W = \vec{N}\cdotp \delta\vec{ \mu}, \label{eq_WeqQV}
\end{equation}
where $W$ is the work per cycle performed by CIFs in the low bias limit and for an infinitesimally slow motion. The vector $\vec{N}$ contains the number of particles pumped per cycle over each conduction channel. This expression, which essentially reflects energy conservation, can be seen as the generalization of similar relations discussed in previous works\cite{AQM13,FBP15,Liliana_onsager} to multi-channel systems with spatially distributed decoherent processes.

The coefficients of the decoherence-induced dissipative forces $\gamma _{\nu\nu ^{\prime }}^{\phi}$, are given by
\begin{equation}
\gamma _{\nu \nu ^{\prime }}^{\phi}= h \left( \frac{\mathrm{d}\vec{n}^{(*)}_{\phi }}{\mathrm{d}x_{\nu }} \right)^T \left( -\mathbb{T}_{\phi \phi }\right)
^{-1}\frac{\mathrm{d}\vec{n}_{\phi }}{\mathrm{d}x_{\nu ^{\prime }}}. \label{eq_gammaphi}
\end{equation}
It can be proved that the matrix $\gamma^\phi$, whose elements are $\gamma_{\nu \nu ^{\prime }}^{\phi}$, is positive definite in our case, spinless particles in the absence of magnetic fields, see Appendix \ref{sec_APossitivity}.

Finally, the third term of Eq. \ref{eq_CIF_dec} is the decoherence-induced fluctuating forces, which is  characterized by its self-correlation
$ \left\langle \xi _{\phi,\nu }(t)\xi _{\phi,\nu ^{\prime
}}(t^{\prime })\right\rangle = D_{\nu \nu ^{\prime }}^{\phi}\delta (t-t')$, where
\begin{eqnarray}
D_{\nu \nu ^{\prime }}^{\phi} & = & 
\frac{h^2}{e^2}
\left( \frac{\mathrm{d}\vec{n}^{(*)}_{\phi }}{\mathrm{d}x_{\nu }} \right)^T
\left( - \mathbb{T}_{\phi \phi }\right)^{-1}	 \times
\nonumber \\
& &
\left\langle
\delta \vec{I}_{\phi }
\otimes
\delta \vec{I}_{\phi }
\right\rangle \left[ \left( - \mathbb{T}_{\phi \phi }\right)^{-1}\right]^T 
\frac{\mathrm{d}\vec{n}^{(*)}_{\phi }}{\mathrm{d}x_{\nu'}}. \label{eq_DFexact}
\end{eqnarray}
Using Eq. \ref{eq_DI} in Eq. \ref{eq_DFexact} and assuming the absence of external magnetic fields, yields
\begin{equation}
D_{\nu \nu ^{\prime }}^{\phi}=2k_{B}T~\gamma _{\nu \nu ^{\prime }}^{\phi}.
\end{equation}
This completes the fluctuation-dissipation relation between $\gamma$ and $\xi$, now including decoherence.

Now, let us neglect fluctuations in the CIFs and currents. This should be equivalent to consider their average value over several realizations of the same experiment.
Then, we can put Eq. \ref{eq_currents_complete} and the right hand side of Eq. \ref{eq_Langevin} in the form
\begin{equation}
\left( \begin{array}{c}
-\overrightarrow{F} \\ 
\overrightarrow{I}%
\end{array} \right)%
=%
\left( \begin{array}{c}
-\overrightarrow{F}^{eq} \\ 
\overrightarrow{I}^{eq}%
\end{array} \right)%
+
\left(\begin{array}{cc}
\mathbb{L}_{11} & \mathbb{L}_{12} \\ 
\mathbb{L}_{21} & \mathbb{L}_{22}
\end{array} \right)
\left(\begin{array}{c}
\overrightarrow{\dot{x}} \\ 
\overrightarrow{\delta \mu }%
\end{array} \right)
.  \label{eq_onsager_matrix}
\end{equation}
We can consider the forces $\overrightarrow{F}$ and the currents $\overrightarrow{I}$
as generalized fluxes and velocities $\overrightarrow{\dot{x}}$ and chemical
potentials $\overrightarrow{\delta \mu }$ as generalized forces in a
thermodynamic sense. The coefficients $\mathbb{L}_{12}$ and $\mathbb{L}_{21}$ can be identified with Onsager's coefficients, which in our system should accomplish $\mathbb{L}_{\nu \alpha}(\vec{B})=-\mathbb{L}_{\alpha \nu}(-\vec{B})$, see Ref. \cite{Liliana_onsager,cohen}.
This can be readily verified in Eqs. \ref{eq_currents_complete} and \ref{eq_NEForces}
\begin{equation}
\left . \frac{\partial F_{\nu}}{\partial \left (\delta \mu _{\alpha} \right )}\right|_{eq,\vec{B}}= - \mathbb{L}_{\nu,\alpha}(\vec{B})=
\mathbb{L}_{\alpha,\nu}(-\vec{B})= \left . \frac{\partial I_{\alpha}}{\partial \dot x_\nu}\right|_{eq,-\vec{B}}.  \label{eq_onsager_element}
\end{equation}

Finally, one can check that the model is consistent with the second law of thermodynamics.
As shown in Ref. \cite{Liliana_entropy_prod} the rate of entropy production $\dot S^{\mathrm{diss}}$ for a driven quantum electronic system is
\begin{equation}
 T\dot S ^{\mathrm{diss}} = -\left ( \vec F - \vec{F}^{eq}\right ) \cdot \vec{\dot x}  + \frac{1}{e} \vec I \cdot \vec{\delta\mu}.
\end{equation}
In our case, this results in
\begin{eqnarray}
T \dot S ^{\mathrm{diss}} & = & \vec{\dot x}^T \left( \gamma^{eq}+\gamma^{\phi} \right )  \vec{\dot x} +
 \vec{\delta\mu}^T \left( -\frac{e}{\hbar} \mathbb T \right) \vec{\delta\mu}  \nonumber \\
 && +\sum_{\alpha \nu}
 \left ( \frac{\mathrm{d}n^{ }_{\alpha}}{\mathrm{d}x_{\nu }} -
 \frac{\mathrm{d}n^{(*)}_{\alpha}}{\mathrm{d}x_{\nu }} \right ) \delta \mu_\alpha \dot x_\nu . \label{eq_Sdiss}
\end{eqnarray}
Using \ref{eq_gamma_simet_frict}, \ref {eq_gammaphi}, and \ref{eq_onsager_element}, together with Appendix \ref{sec_APossitivity} one can readily check that for the case of interest, spinless particles in the absence of magnetic fields, the rate of entropy production $\dot S ^{\mathrm{diss}}$ is always positive.

\subsection{Thouless quantum motor} \label{subsec_System}

\begin{figure}[tbp]
\begin{center}
\includegraphics[width=1.5 in]{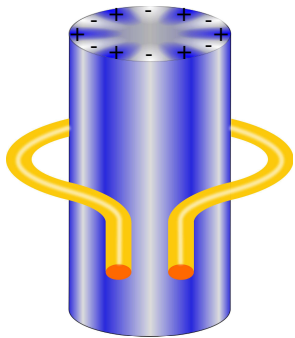}
\end{center}
\caption{Scheme of a possible Thouless motor made out of charges periodically arranged on the surface of a rotational piece and interacting with a wire coiled around it. Rotation of the cylinder changes the potential sensed by the electrons in the wire.}
\label{fig_Thouless}
\end{figure}

\textit{General model - } A Thouless quantum motor is basically the reverse of a Thouless pump. In principle, this can be accomplished in several ways. One possible realization of it is schematized in Fig. \ref{fig_Thouless}. In our proposal the Thouless motor consists of just a conducting wire, which could be a conducting polymer for example, coiled around a rotational piece with periodically arranged charges on its surface.
For simplicity, in the following we will consider a single-channel wire where the potential of the conduction electrons traversing the wire is given by 
\begin{equation}
U(x)=2 \Delta \cos (2\pi x/\lambda -\theta).
\end{equation}
Here $\Delta $ is the coupling strength with the classical coordinate, $\lambda $ is the period and $\theta$ varies according to the angular position of the rotor. The mechanical rotation is described uniquely by the coordinate $\theta$, which we assume slow with respect to the dwell time of the electrons inside the wire. The dynamics of the rotor will be described by the angular Langevin equation
\begin{equation}
M_\theta \ddot{\theta} = F_\theta -\gamma_\theta \dot{\theta} +\xi_\theta. \label{eq_LangavineTheta}
\end{equation}
Here $M_\theta$ is the moment of inertia of the rotor, $F_\theta$ is the mean adiabatic reaction force, which is a torque in the present case, $\gamma_\theta \dot{\theta}$ is the friction force where $\gamma_\theta$ is the friction coefficient in units of ``$\mathrm{Energy} \times \mathrm{time} / \mathrm{radians}^2$'', and $\xi_\theta$ is the noise component of the CIFs, also a torque.

\textit{Tight-binding model - } To calculate all the terms of the CIFs, we will resort to a tight-binding model. The electronic Hamiltonian of the wire is
\begin{equation}
\hat{H}_{S}=\sum\limits_{n=1}^{N}\left\{ E_{n}\hat{c}_{n}^{\dagger }\hat{c}%
_{n}^{{}}-V_0 \left[ \hat{c}_{n}^{\dagger }\hat{c}_{n-1}^{{}}+\hat{c}%
_{n-1}^{\dagger }\hat{c}_{n}^{{}}\right] \right\} , \label{eq_Hs}
\end{equation}
where $\hat{c}_{n}^{\dagger}$ and $\hat{c}_n$ are the creation and annihilator operators at site $n$, $V_0$ is the hopping, and $E_{n}=U(x)$ is the site energy, being $x=na$ with $a$ the lattice constant and $n$ the site index.
We will include the leads through self energies, $\Sigma$, acting on the first and last sites, $1$ and $M$, where 
\begin{equation}  \label{Dyson}
\Sigma (\varepsilon ) = \lim_{\eta \rightarrow 0^+} \frac{\varepsilon+\mathrm{i}\eta }{2}
-\mathrm{sgn}(\varepsilon)\sqrt{\left( \frac{\varepsilon+\mathrm{i}\eta }{2}\right)
^{2}-V_0^{2}}.
\end{equation}
Then, the effective Hamiltonian results in
\begin{equation}
\hat{H}_{eff}=\hat{H}_{S}+\Sigma (\varepsilon )[\hat{c}_{1}^{\dagger }\hat{c}^{\ }_{1}+\hat{c}_{M}^{\dagger }\hat{c}^{\ }_{M}].
\end{equation}

Note that this Hamiltonian presents a jump in the site energies $E_{n}$ between the leads and the system.
Such an abrupt change of the potential energy can cause the scattering of the incoming or outgoing wave functions which in turn may affect CIFs.
This effect, not previously discussed for AQMs, should be present in almost any device acting as an AQM and, as we will discuss in the next section, it can dramatically affect the dynamics of the system.

\begin{figure}[tbp]
\begin{center}
\includegraphics[width=1.8 in]{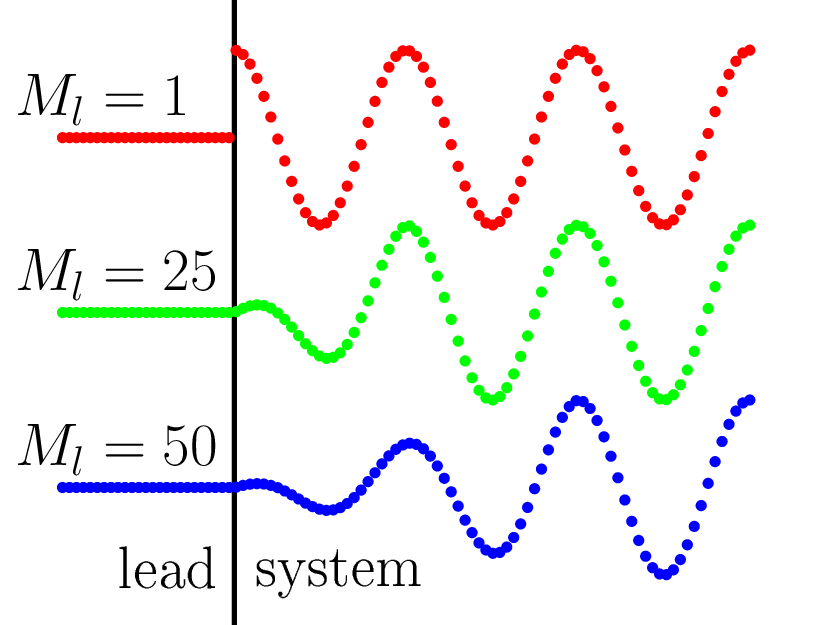}
\end{center}
\caption{Scheme showing how the site energies change for different system-lead transitions. See section \ref{subsec_System}, \textit{Tight-binding model}.}
\label{fig_edges}
\end{figure}

In order to account for different system-lead junctions we use a linear function $f(n)$ that multiplies the site energies $E_n$ of the $M_l$ firsts and lasts sites. The term ``$M_l$'' in the figures stands for $f(x)= x /M_l$, where $x=a n$ or $x=a (M_l-n)$ for the first or last sites, respectively. In this way the larger the $M_l$ the smoother the lead-system transition; $M_l=1$ correspond to the ``abrupt'' transition, see Fig. \ref{fig_edges}.

Decoherence is included in this model by adding fictitious leads to every tight-binding site with a self-energy given by $\Sigma_\phi =-i\Gamma_\phi$. 
The final effective Hamiltonian including decoherence yields,
\begin{equation}
\hat{\widetilde{H}}_{eff}=\hat{H}_{S}+\Sigma (\varepsilon )[\hat{c}_{1}^{\dagger }\hat{c}^{\ }_{1}+\hat{c}_{M}^{\dagger }\hat{c}^{\ }_{M}] + \sum_{n=1}^N \Sigma_\phi \hat{c}_{n}^{\dagger }\hat{c}^{\ }_{n}.
\end{equation}
The elements of the $\mathbb{T}$ matrices in 
Eqs. \ref{eq_current_block}, \ref{eq_Ueq}, \ref{eq_NEForces}, and \ref{eq_gammaphi} are then calculated using the Fisher and Lee formula, see Refs. \cite{CFBP14} and \cite{PM01}.

\textit{Ideal model - } The Thouless motor has been studied previously using what we call here the "ideal model".\cite{AQM13} In this model the Hamiltonian is linearized with respect to the momentum of the electrons, resulting in an effective Hamiltonian with
counter propagating linear channels and backscattering due to the periodic potential,
\begin{equation}
\hat H = v_F  p \sigma^z + \Delta \left (\sigma^x \cos\theta + \sigma^y \sin \theta \right ) \Theta(L/2-|x|).
\label{eq_Hideal} 
\end{equation}
Here $L$ is the length of the system, $\Theta$ is the step function, the momenta are measured from $\pm k_0$ and energies from $\hbar^2 k_0^2/2m$ with $k_0=\pi/\lambda$, the ($\sigma^i$)s denote the Pauli matrices in the space of the counter propagating channels, and the real electron spin is not included for simplicity. Using this Hamiltonian and assuming a perfect system-lead matching one can solve analytically the scattering matrix and its derivatives for the Thouless motor. In the limit $L \rightarrow \infty$ and for electron's energies within the band gap, one finds that the charge pumped per cycle and the efficiency are one while the equilibrium forces and the transmittance are zero.\cite{AQM13} 

\textit{Work and efficiency -} To avoid confusions we will define some magnitudes in the stationary regime that will be discussed in the following sections. The adiabatic work of CIFs, $\lim_{\dot\theta\rightarrow 0} W=W^{(a)}(\theta)$, is
\begin{equation}
W^{(a)}(\theta)=\int_0^\theta F(\theta') d\theta'.
\end{equation}
The total work per cycle $W^{\mathrm{(total)}}$ is
\begin{equation}
W^{\mathrm{(total)}}=W^{(a)} - W^{\mathrm{(load)}} - \int_0^\tau \gamma(t') \dot \theta^2 dt', \label{eq_Wtotal}
\end{equation}
where $W^{\mathrm{(load)}}$ is the work exerted by some external force against which one wants to move the system, and $\tau$ is the period of the motor.
The output power per cycle $\dot W^{\mathrm{(load)}}$ can be obtained from the steady state condition $W^{\mathrm{(total)}}=0$, while the input power is simply the total current times the applied voltage. Then, the thermodynamic efficiency, calculated as the average output power per cycle over the average input power per cycle, results in
\begin{equation}
\eta_{TD} = \frac{ N \delta\mu - \int_0^\tau \gamma(t') \dot \theta^2 dt'} 
{\tau \left< T_{LR}\right> \delta\mu^2+ N \delta\mu}. \label{eq_etaVsmall}
\end{equation}
Here, we have used Eq. \ref{eq_WeqQV} and assumed a two leads system with an arbitrary number of conduction channels. $\left< T_{LR}\right>$ is the sum of the average transmittances (averaged over time) that connects all channels of one lead with all channels of the other, $N$ is the total charge pumped per cycle to one lead through any channel of it, and $\delta\mu=|\mu_L-\mu_R|$ is the chemical potential difference among the leads.

The evaluation of $\eta_{TD}$ requires the integration of the equation of motion of the problem. However, this can be circumvented by assuming a terminal velocity approximately constant. Then, the thermodynamic efficiency yields
\begin{equation}
\eta_{TD} \approx \frac{ N - 4 \pi^2 \left< \gamma\right>/\left ( \tau \delta\mu \right ) } 
{\tau \left< T_{LR}\right> \delta\mu+ N }, \label{eq_etaVsmallThetaConst}
\end{equation}
where $\left< \gamma\right>$ is the average friction coefficient, and the value of $\tau$ is left as a free parameter which depends ultimately on $W^{\mathrm{(load)}}$.
If we assume $F^{\mathrm{(load)}}(\theta)$ independent of $\dot\theta$, the period is given by
\begin{equation}
\tau \approx 4 \pi^2 \frac{\left< \gamma\right> } { N \delta\mu-W^{\mathrm{(load)}}}. \label{eq_tauWconst}
\end{equation}
Alternatively, if we assume $F^{\mathrm{(load)}}(\theta)$ proportional to $\dot\theta$, one can write
\begin{equation}
\tau \approx 4 \pi^2   \frac{ \left< \gamma+\gamma^{\mathrm{(load)}}  \right>}
{ N \delta\mu }. \label{eq_tauWfricc}
\end{equation}

As shown in Appendix B of Ref. \cite{FBP15} and in panel C of Fig. \ref{fig_map}, taking $\dot\theta$ constant in the steady state results in a good approximation for a large momentum of inertia. This approximation will be used in Figs. \ref{fig_Eff},  \ref{fig_EffPowervsTau}, and \ref{fig_Lygamma} of the next section.

\section{Results} \label{sec_Results}
The Thouless motor has shown to be an optimal candidate for the realization of an AQM since its maximum theoretical efficiency is one.\cite{AQM13}
In this section we will evaluate its performance under nonideal conditions such as nonlinear dispersion relations, nonideal system-lead matchings, external friction sources, and decoherence of the electrons responsible for the motor's motion.

\subsection{Deviations from the ideal model} \label{subsec_Deviations}

\begin{figure}[tbp]
\begin{center}
\includegraphics[width=2.8in]{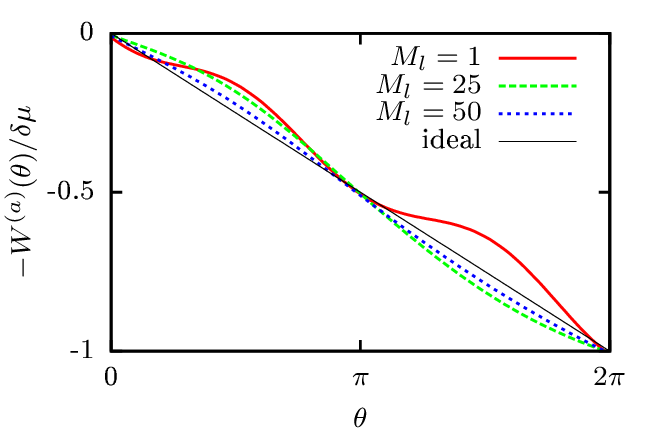}
\end{center}
\caption{Effect of the border conditions on the adiabatic work $W^{(a)}$. The line labeled `ideal' correspond to the expected result from the ideal model.}
\label{fig_WvsBC}
\end{figure}

\begin{figure}[tbp]
\begin{center}
\includegraphics[width=2.5 in]{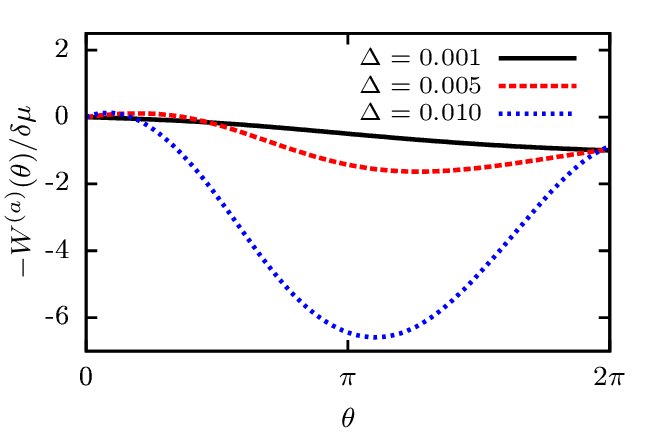}
\end{center}
\caption{Effect of $\Delta$ (in units of $V_0$) on the adiabatic work $W^{(a)}$.}
\label{fig_WvsDelta}
\end{figure}

The proposal of the Thouless motors given in Ref. \cite{AQM13} involves several simplifications to obtain an analytical solution of the CIFs.
One of these simplifications consists of neglecting the scattering of electrons entering the system.
However, the potential profile inside the system is periodic, and it changes with the coordinate of the mechanical degree of freedom while the potential profile in the leads is constant and fixed. Then, there must be some $\theta$-dependent scattering of electrons and consequently, CIFs must be affected by the details of the junctions. It is worth mentioning that this effect prevails at equilibrium ($\delta \mu = 0$) and is a consequence of the quantum nature of the particles driving the motion of the system.
Figure \ref{fig_WvsBC} shows an example of the emergence of corrections to the CIFs in the Thouless motor as a consequence of the ``edge mismatching''.
There, the work performed as a function of the coordinate shows oscillations that decrease with the softening of the system-lead transition.
The strength of this edge effect will depend, of course, on the details of the system-lead interface but also on the coupling strength $\Delta$ as can be seen in Fig. \ref{fig_WvsDelta}.
According to this figure one way to reduce this edge mismatching is to reduce the coupling strength $\Delta$. However, this can not be done without consequences as it is always necessary a gap that makes the transmittance zero and drives the efficiency towards one. For this reason a smaller value of $\Delta$ will require a larger wire, which could bring its own consequences such as a larger decoherence rate.

\begin{figure}[tbp]
\begin{center}
\includegraphics[width=2.5in]{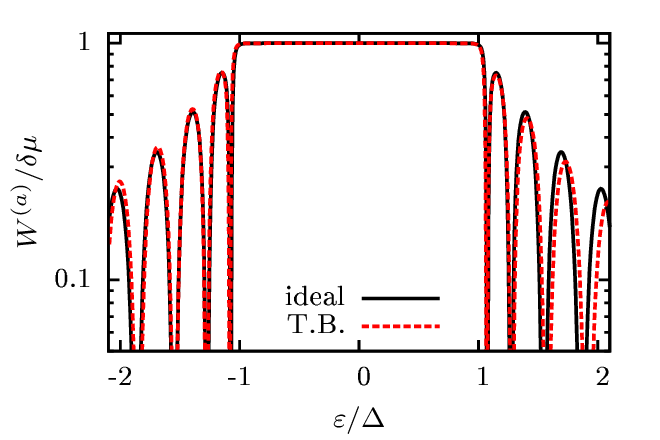}
\end{center}
\caption{Comparison of the adiabatic work $W^{(a)}$ per cycle calculated from the tight-binding model (T.B.) and the ideal model (ideal).
Fermi energy $\varepsilon$ measured from the center of the band gap.}
\label{fig_WvsEta}
\end{figure}

Taking into account that the ideal model of the Thouless motor is based on a linearized Hamiltonian, some deviations are expected from the results with respect to the energy of the electrons, especially at energies far from the center of the band gap. However, as can be seen in Fig. \ref{fig_WvsEta} the differences are not significant, at least at small values of the coupling strength $\Delta$.

\subsection{Dynamics} \label{subsec_Dynamics}

\begin{figure}[tbp]
\begin{center}
\includegraphics[width=3.3in]{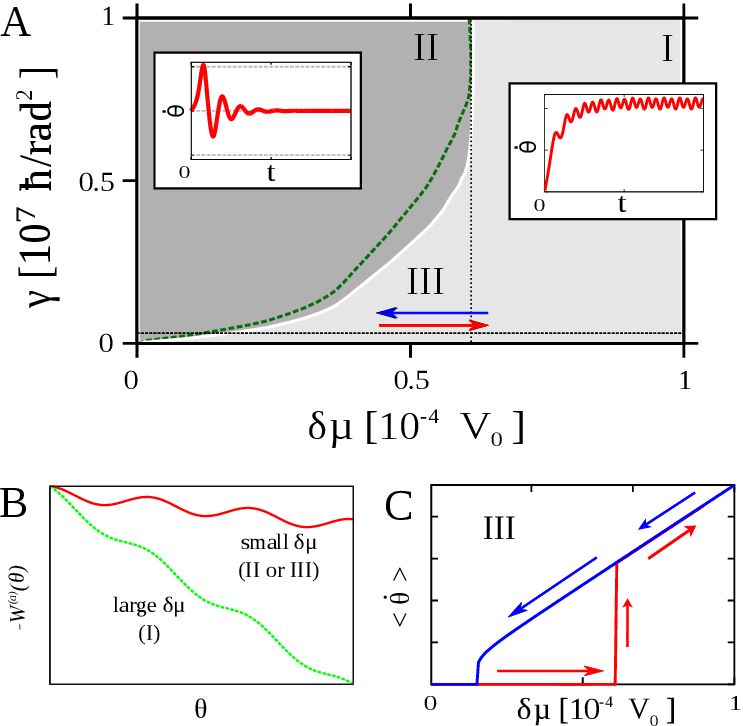} 
\end{center}
\caption{$\textbf{A - }$Phase diagram of different steady state dynamics as a function of $\delta\mu$ and $\gamma$, assumed constant for simplicity, for $W^{\mathrm{(load)}}=0$.
Insets of the regions I and II show typical dynamics for these regions ($\dot\theta$ vs time). The typical dynamics of the region III is shown in panel $C$.
$\textbf{B - }$ Scheme of how $W^{(a)}(\theta)$ changes with $\delta\mu$.
$\textbf{C - }$ Average terminal velocity $<\dot\theta>$ versus $\delta \mu$ for a typical hysteresis cycle where the dynamics can fall into the cases shown in the insets of the regions I or II depending on the initial conditions.
Panels C was calculated with the value of $\gamma$ shown as a horizontal line in panel $A$.
The vertical line in $A$ corresponds to the value of $\delta \mu$ to which $W^{(a)}(\theta)$ vs $\theta$ does not show a minimum.
Dashed green line in panel A showed the transition between the regions II and III but for twice the moment of inertia.}
\label{fig_map}
\end{figure}

One of the key parameters that affect the dynamics of the studied system is the friction coefficient $\gamma$.
In realistic situations dissipation of the mechanical energy can arise from different sources not only from the CIFs. Therefore, there is a minimal value of $\gamma$, that arising from CIFs, but not a maximum one.
In order to study the effect of different parameters on the dynamics, we solved the equation of motion of Thouless motors using Eq. \ref{eq_LangavineTheta} for a wide range of different conditions. The results are summarized in the phase diagram shown in Fig. \ref{fig_map}.
In the calculations we took for simplicity $\gamma$ as constant, an arbitrary moment of inertia $M_\theta=10^{18} \hbar^2/(V_0\mathrm{rad}^2)$, and the low temperature limit $T \approx 0$. In Appendix \ref{sec_AFeasibility} we discuss about the parameters used in this section and the role of $M_\theta$ on the dynamics.

As previously shown, the system-lead mismatching introduces corrections to the CIFs, and one of the consequences is the oscillatory behavior of $W^{(a)}(\theta)$.
This effect induces, under the appropriate conditions, a dependence of the final state on the initial conditions.
This can be understood considering the following example.
Suppose $W^{(\mathrm{total})}$ possesses a minimum, as the case labeled ``small $\delta\mu$'' in panel B of Fig. \ref{fig_map}. Then, let us assume the motion starts from the first maximum to the left of that minimum with a temperature close to zero and with a small velocity pointing to the right.
Additionally, let us assume the energy dissipated as friction is lower than the energy gained by reaching the first maximum to the right. Under this condition, the system will keep moving to the right accelerating its motion until the energy dissipated becomes so large that $W^{(\mathrm{total})}=0$.
This dynamics will look much like that shown in the inset of the region I of Fig. \ref{fig_map}.
If we start from the minimum under the same condition, the system will follow a damped oscillatory dynamics as that shown in the inset of the region II of Fig. \ref{fig_map}.
Both situations are present in the region III of Fig. \ref{fig_map}. There, if the rotor is still, it will remain still, but if it is already rotating, it will continue to turn.

The above mentioned effect is not present at larger $\delta \mu$, where $W^{(a)}(\theta)$ does not present a minimum.
In this case, independently of the initial condition, the system will move until the stationary condition is reached, $W^{(\mathrm{total})}=0$, at a high enough average terminal velocity.
This situation corresponds to the region I of the phase diagram of Fig. \ref{fig_map} and its typical dynamics looks like that of its correspondent inset.
The other possibility is that $W^{(a)}(\theta)$ does present a minimum, but $\gamma$ is so large that the energy gained after one period does not compensate the energy lost through friction, even if the initial velocity is infinitesimally small.
This situation corresponds to the region II of the phase diagram of panel A of Fig. \ref{fig_map}. Its typical dynamics looks like that of its correspondent inset.
Panel C of Fig. \ref{fig_map} shows the average terminal velocity resulting from different dynamics calculated by varying the applied voltage progressively. We first increase the voltages from zero (lower red arrow) and then, starting from a high voltage, we decrease it (upper blue arrow). Note the hysteresis cycle that occurs in the region III and the linear dependence of $<\dot\theta>$ on $\delta\mu$ which is consistent with $\dot\theta(\theta)$ approximately constant, see Eq. \ref{eq_tauWconst}.

The effect of a load coupled to the motor will be like that of an additional friction or a constant force. The first case occurs, for instance, when the motor has to drag a mass against an external friction source. The second one occurs, for example, when the motor is lifting up a mass against gravity. In both cases, the behavior of the system will be much like that described above, but with either a corrected value of $\gamma$ or a corrected value of $\delta\mu$, respectively.

In this section, our intention was to highlight the importance of the boundary induced CIFs in the dynamics.
For this reason, we only discussed the zero temperature limit of the dynamics, where the difference between dynamical regimes is clearer.
At finite temperatures, crossing a barrier becomes just a matter of probabilities or waiting enough time.
Under this condition, the discussed behaviors are still present but blurred by temperature, and the transitions from one regime to the other are smoother. If the temperature is high enough, the dynamics looks like that of a random walk where even backward motion is allowed and the average velocity results from the difference in the probabilities of moving in one or the other direction, (see Ref. \cite{Hanggi_brownian} and references therein).

\subsection{Effect of decoherence} \label{subsec_Decoherence}

In an ideal Thouless motor, all electrons incoming to the wire with energy within a certain range are reflected.
Conservation of momentum imposes that electron back-scattering should move the rotor.
As transmittance is close to zero within the band gap, $I^{total} \approx I^{pump}$, Onsager's relations, Eq. \ref{eq_onsager_element}, imply that all electrons passing through the system contribute to the work per cycle.
The result of this, is that the efficiency goes towards one in the limit of small velocities, see Eq. \ref{eq_etaVsmall} .
Since the suppression of the transmittance is a consequence of the quantum nature of the electrons, its is natural to wonder about the performance of the motor under dephasing environments.

Let us assume the rate of decoherence, $2 \Gamma _{\phi }/\hbar = 1/\tau_\phi$, is constant along the wire but negligible within the leads.
Then, electrons crossing a wire of total length $L$ and decoherent length $L_{\phi}=2 v_f \Gamma _{\phi }/\hbar$ have a probability $P \propto e^{-L/L_\phi}$ of traversing the system without undergoing a decoherent event.\cite{DP90}
If the total length of the system $L$ is close to $L_{\phi}$ then it will be necessary to include the effect of decoherence on CIFs according to the theory developed in section \ref{sec_Theory}.
\begin{figure}
\begin{center}
\includegraphics[width=2.5in, trim= 0.0in 0.0in 0.0in 0.0in, clip=true ]{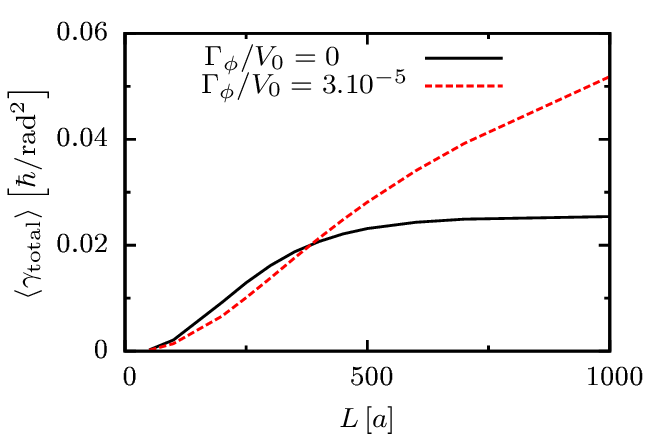}
\end{center}
\caption{
Effect of decoherence and the system's length $L$ on the averaged total friction coefficient $\left < \gamma_{\mathrm{total}} \right >$. Here, $\gamma$ stands for the current-induced friction coefficient.
}
\label{fig_GammavsTheta}
\end{figure}
\begin{figure}[tbp]
\begin{center}
\includegraphics[width=2.5in, trim= 1.0in 0.3in 0.5in 0.6in, clip=true]{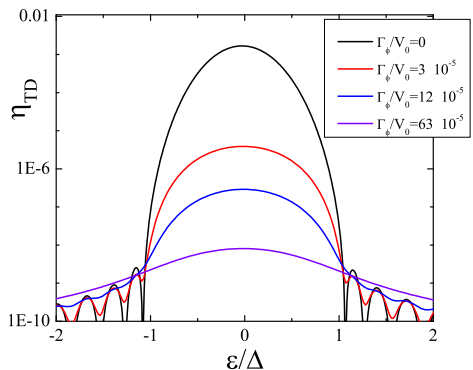}
\end{center}
\caption{
Effect of decoherence on the thermodynamic efficiency $\eta_{TD}$ for different values of the Fermi energy, measured from the center of the band gap. We assumed an external dissipative mechanism with $\gamma_{\mathrm{ext}}=10^8 \hbar/\mathrm{rad}^2$.
Each curve was calculated using the value of $\tau$ that maximizes $\eta_{TD}$ for $\varepsilon=0$, see Eq. 15 of Ref. \cite{AQM13}.
}
\label{fig_Eff}
\end{figure}

Figure \ref{fig_GammavsTheta} shows the effect of decoherence and the length of the system on the total current-induced friction coefficient $\gamma_{\mathrm{total}}=\gamma^{eq}+\gamma^{\phi}$.
First, note that there is a saturation of $\gamma_{\mathrm{total}}$ at large $L$.
This is reasonable since the Fermi energy is within the band gap and then electrons possess a penetration depth beyond which the system is not further explored.
This penetration depth is increased by decoherence, and this, in turn, increases the value of $L$ at which $\gamma_{\mathrm{total}}$ saturates.
The other interesting feature to be noticed is the change in the behavior of $\gamma_{\mathrm{total}}$ with or without decoherence for systems of different lengths.
Note that for short systems, decoherence always reduces the friction, while for long systems it is the opposite.
This can be understood as two competing effects of decoherence on $\gamma$.
First, the type of decoherence described by our model does preserve energy but not momentum, which in turn induces a random-walk-like dynamics of electrons, in a semiclassical picture.
This phenomenon should increase the noise and then $\gamma_{\mathrm{total}}$ with $\Gamma_\phi$.
Second, for small systems, adding reservoirs or increasing the connection to them tends to reduce the noise in the currents, which on the other hand is the source of noise of CIFs.\cite{beenakker1992suppression,Liu_decoherence_noise}
Then it is expected for small systems a decrease of $\gamma_{\mathrm{total}}$ with $\Gamma_\phi$. As can be seen in Fig. \ref{fig_GammavsTheta}, the first effect dominates at large ($L$)s while the second one dominates at small ($L$)s.

\begin{figure}[tbp]
\begin{center}
\includegraphics[width=2.3in, trim= 1.0in 0.1in 0.7in 0.8in, clip=true]{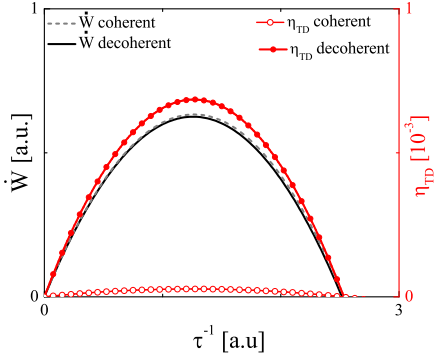}
\end{center}
\caption{Effect of decoherence on the thermodynamic efficiency $\eta_{TD}$ and the output power ${\dot W}^{\mathrm{(load)}}$ (in arbitrary units) for different operational frequencies of the motor $\tau^{-1}$ (in arbitrary units). We assumed an external dissipative mechanism with $\gamma_{\mathrm{ext}}=10^8 \hbar/\mathrm{rad}^2$.}
\label{fig_EffPowervsTau}
\end{figure}

\begin{figure}[tbp]
\begin{center}
\includegraphics[width=2.3in, trim= 0.1in 0.0in 6.0in 0.1in, clip=true]{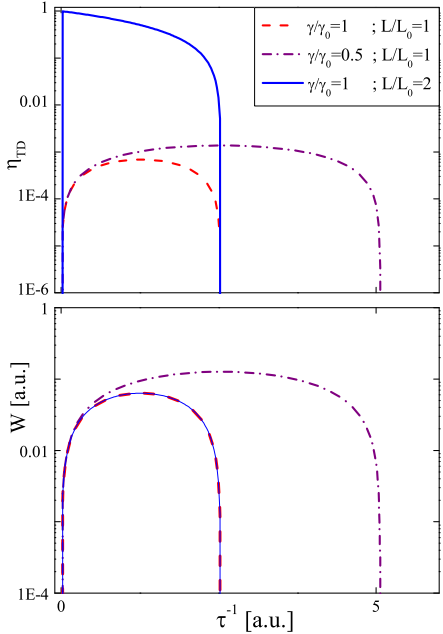}
\end{center}
\caption{Effect of $\gamma$ and the system's length $L$ on the thermodynamic efficiency $\eta_{TD}$ and the output power $\dot W$ (in arbitrary units). $L_0=2000a$ and $\gamma_0=10^8 \hbar/\mathrm{rad}^2$.}
\label{fig_Lygamma}
\end{figure}

Let us note that in Fig. \ref{fig_GammavsTheta}, the analyzed friction coefficient stands for the minimum possible energy dissipation that the system may suffer, i. e. when any external source of friction are suppressed.
In Figs. \ref{fig_Eff}, \ref{fig_EffPowervsTau}, and \ref{fig_Lygamma}, we assume that the current-induced friction is much lower than external sources of friction. The value used in the figures of the external friction coefficient $\gamma_{\mathrm{ext}}$, was taken such as the terminal velocity falls close to that of experimental nanomotors, see Appendix \ref{sec_AFeasibility}.
In Fig. \ref{fig_Lygamma} we analyze the effect of $\gamma$ and the system's length on the efficiency and the output power.

In Fig. \ref{fig_Eff} we plot the efficiency for different values of $\Gamma_\phi/V_0$.
For the pure coherent case, $\Gamma_\phi/V_0 = 0$, one can see a strong oscillatory behavior of the efficiency outside the band gap. This feature was already present in the analytic solution of the model proposed in Ref. \cite{AQM13}.
They are just a consequence of resonances of the reflection coefficients for certain values of the Fermi energy and the length of the system, similarly to the Fabry-Perot resonances.
As expected, decoherence has a profound influence on the efficiency.
For a decoherence rate of $3\ 10^{-5}\ V_0$, the efficiency smoothes notably its resonances and drops its maximum value in two orders of magnitude.
To put this value of $\Gamma_\phi$ into context, a tight-binding chain of the same length but without the periodic potential gives $L/L_{\phi}=0.5$. This implies that almost half of the electrons suffer a decoherent event while passing through the sample.
It is interesting that some memory of the quantum nature of the Thouless motor remains even at high decoherence rates.
The effect of the band gap on the efficiency is still noticeable even for $\Gamma_\phi/V_0=63\ 10^{-5}$.
This decoherence rate, for the same system but without the periodic potential,
gives $L/L_{\phi}=10$, which implies that only 1 out of 22000 electrons pass the system without suffering a decoherent event.

It is interesting to study the effect of the period of the motor $\tau$ on the efficiency $\eta_{TD}$ and the output power $\dot W$. This is done in Fig. \ref{fig_EffPowervsTau}, where we compare the coherent and the decoherent cases. From the figures, it is clear that $\eta_{TD}$ results much more sensitive to decoherence than $\dot W$. This shows that the decrease in the efficiency due to decoherence is just a consequence of the increase of the transmittance and not of the decrease in $W$, see Eq. \ref{eq_etaVsmall}.

Before discussing the last figure, first note that the values of $\eta_{TD}$ in Figs. \ref{fig_Eff} and \ref{fig_EffPowervsTau} are far from the optimal value predicted by the ideal model, $\eta_{TD}=1$.\cite{AQM13}. This is only a consequence of the transmittance not being small enough for the value of $\gamma$ used. In Eq. \ref{eq_etaVsmall} one can check that for $T_{LR}$ strictly zero, one gets $\eta_{TD}=1$ in the limit of $\tau \rightarrow \infty$, independently of the value of $\gamma$. 
However, for finite $T_{LR}$ the larger the value of $\gamma$, the smaller the value of $\eta_{TD}$. Regretfully, due to numerical limitations, we were not able to evaluate the effect of decoherence on $\eta_{TD}$ for larger systems, which implies smaller $T_{LR}$. However, in Fig. \ref{fig_Lygamma} we show the effect of $L$ and $\gamma$ on $\eta_{TD}$ and $\dot W$ for the coherent case.
There one can see, that doubling the length of the system has a strong influence on $\eta_{TD}$,  for $\tau^{-1} \approx 0$ $\eta_{TD} \approx 1$.
Decreasing $\gamma$ also affects favorably $\eta_{TD}$, but its effect is not too strong.
Finally, note that the conditions that maximize $\eta_{TD}$ and $\dot W$ do not necessarily coincide, especially for highly efficient motors. This is reasonable after analyzing Eq. \ref{eq_etaVsmall}. There, it is clear that for $I^{bias}=0$, $\tau \rightarrow \infty$ maximize $\eta_{TD}$ by minimizing the energy dissipated as friction, while this condition makes $<\dot W >= W/\tau \rightarrow 0$.
The consequence of this, is that one has to choose between maximizing the output power or the efficiency in a Thouless motor.

\section{Conclusions} \label{sec_conclusions}

We have extended our previous theory of decoherence in CIFs to account for spatially distributed decoherent processes providing analytical expressions for the CIFs, friction coefficients and the self-correlation functions of the forces.
We have proved that our model is thermodynamically consistent, fulfilling fluctuation-dissipation theorem, Onsager's reciprocity relations, and the first and the second laws of thermodynamics.

We have confirmed that decoherence drastically reduces the efficiency of the motor mainly due to the increase in conductance, while its effect on the output power is not too important.
The effect of decoherence on the current-induced friction depends on the length of the system, reducing the friction for short systems while increasing it for long ones.

We have found that the conditions that maximize the efficiency do not necessarily coincide with those that maximize the output power. This could have important consequences for the implementation of the motor.

We have shown that the system-lead mismatching produces conservative forces that can dramatically alter the dynamics of motors.
At high dissipation, these forces set a minimum voltage that allows the operation of the motor. At low dissipation, the boundary-induced forces cause hysteresis with two limiting voltages that switch on-off the movement of the motor depending on its previous history.

\section{Acknowledgement.}

This work was supported by CONICET (Consejo Nacional de Investigaciones Cient\'ificas y T\'ecnicas), SECYT-UNC (Secretaria de Ciencia y Tecnología - Universidad Nacional de C\'ordoba) and ANPCyT (Agencia Nacional de Promoci\'on Cient\'ifica  Tecnol\'ogica).

\appendix

\section{Positivity of $\mathbb{T}$, $\gamma^{eq}$ and $\gamma^{\phi}$ matrices} \label{sec_APossitivity}

Gauge invariance of the bias current implies that a constant shift of the chemical potentials must not yield additional currents through the system. This condition is encoded in the relation $\sum _{\beta \neq \alpha} T_{\alpha \beta}=\sum _{\beta \neq \alpha} T_{\beta \alpha}$ and forces the definition $T_{\alpha \alpha}=-\sum _{\beta \neq \alpha} T_{\alpha \beta}$.
This is the central property that in the end guarantees the positivity of the $\gamma^{\phi}$ and $\mathbb T$ matrices and in turn the positivity of the entropy production with or without decoherence, Eq. \ref{eq_Sdiss}.

The Gershgorin's circle theorem \cite{Gershgorin} applied to our problem, $\mathbb{T}$ symmetric, establishes that
\begin{equation*}
\lambda \in \bigcup_\alpha  D_{\alpha}\mathrm{ \ where \ }D_{\alpha}=\left\{ x \in \mathbb{R}:\left\vert x- T_{\alpha \alpha} \right\vert \leq \sum_{\beta \neq \alpha} 
T_{\alpha \beta} \right\},
\end{equation*}
where $\lambda$ is an eigenvalue of $\mathbb{T}$.
The above implies that all $x$ in the interval $D_{\alpha}$ satisfies
\begin{equation*}
-2<T_{\alpha \alpha }-\sum_{\beta \neq \alpha }T_{\alpha \beta }\leq ~x~\leq
T_{\alpha \alpha }+\sum_{\beta \neq \alpha }T_{\alpha \beta }\le 0,
\end{equation*}%
Thus, in particular, 
\begin{equation*}
 ~x~\le 0.
\end{equation*}
The union of all sets $D_{\alpha}$ does not contain positive values, and then $(-\mathbb{T})$ is a positive semi-definite matrix.

The positivity of the matrix $\gamma^{\phi}$ depends on that of the matrix $\left(-\mathbb{T}_{\phi \phi }\right) ^{-1}$. 
We start by first noticing that $\mathbb{T}_{\phi \phi }$ and $\mathbb{T}^{-1}_{\phi \phi }$ are symmetric for the case of interest. 
Then, the diagonal elements of $\mathbb{T}_{\phi \phi }$ satisfy the following condition
\begin{equation}
\left | \left [\mathbb{T}_{\phi \phi}\right]_{\alpha \alpha } \right | 
=
\sum_{\beta \neq \alpha, \beta \in \ell,\phi} T_{\alpha \beta }
> \sum_{\beta \neq \alpha, \beta \in \phi} T_{\alpha \beta }
\end{equation}
where $T_{\alpha \alpha } < 0,$ and $T_{\alpha \beta }> 0$ for $\alpha \neq \beta$.
Unlike the previous case with the matrix $\mathbb{T}$, the diagonal elements of $\mathbb{T}_{\phi \phi}$
are strictly greater than $\sum_{\beta \neq \alpha, \beta \in \phi}T_{\alpha \beta }$.
By applying the Gershgorin circle's theorem to this case, we conclude that all $\lambda^\phi$'s, eigenvalues
of $\mathbb{T}_{\phi \phi }$, are strictly negative. 
This implies that $\left(-\mathbb{T}_{\phi \phi }\right) ^{-1}$ is positive definite, which ensures the positivity of the matrix $\gamma^\phi$. 

Given the unitarity of the $S$ matrix, which is a consequence of particle conservation, one can readily prove that $[S^{\dagger} \partial S]_{\alpha \beta}=-[S^{\dagger} \partial S]^*_{\beta \alpha}$. Then, using the reciprocal relation for the case of interest in this work, $S_{\alpha \beta}=S_{\beta \alpha}$, and the fact that $-\partial f /\partial \varepsilon \geq 0$ one arrives to
\begin{equation}
\gamma^{eq}_{\nu,\nu} =
\frac{1}{2}\int
\frac{ d \varepsilon }{2\pi }
\left (
-\frac{\partial f }{\partial \varepsilon }
\right )
\sum_{\alpha,\beta}
\left |S^{\dagger }\frac{\partial S}{\partial x_\nu} \right|^2_{\alpha,\beta}
\geq 0                           
\end{equation}
Positivity of $\vec{\dot x}^T \gamma^{eq} \vec{\dot x}$ is obvious by considering a change of basis where $\gamma^{eq}$ is diagonal.

Combining the three previous results we conclude that the rate of entropy production given by Eq. \ref{eq_Sdiss} is always greater or equal to zero, as required by Thermodynamics.

\section{Parameters used and feasibility} \label{sec_AFeasibility}

In all the figures shown in body text, the total size of the system $L$ considered is 2000 sites (except in Fig. \ref{fig_Lygamma}), the period $\lambda$ is 25 sites, the coupling strength $\Delta$ is $0.001 V_0$ (except in Fig. \ref{fig_WvsDelta}), the value of $M_l$ used is 25 (except in Fig. \ref{fig_WvsBC}), the Fermi energy is taken at the center of the gap (except in Figs. \ref{fig_WvsEta} and \ref{fig_Eff}), and the value of $\delta\mu$ is $10^{-4}$ (except in Fig. \ref{fig_map}).

Although the aim of this work is to study general characteristics of the Thouless motors, we consider important to add a brief discussion about the relation between the parameters used and possible experimental scenarios.
For this reason, let us consider a concrete example, a Thouless motor made of some conducting polymer rolled around a cylinder made of $\mathrm{SiO}_2$. With that in mind, we will take the coupling of the tight-binding model $V_0$ approximately equals to that of a $\pi-\pi$ bond, $V_0=-3.6 \mathrm{eV}$ \cite{CBMP10}. Considering the separation between neighboring carbon atoms in a double bond, we will take the lattice constant $a$ as $a=0.14 nm$. Then, the total length of the conducting wire results in $280nm$.

The moment of inertia used in Fig. \ref{fig_map}, $M_\theta=10^{18} \hbar^2/ V_0$, is roughly that of a cylinder of radius $r=65nm$ and height $d=4 r$ made of $\mathrm{SiO}_2$. Anyway, the particular value of $M_\theta$ only determines the scale of the y-axis in the figure. Note, that Eq. \ref{eq_LangavineTheta} can be rewritten as 
\begin{equation}
 \frac{\partial^2 \theta}{\partial t'^2} = F_\theta - \frac{\gamma_\theta}{\sqrt{M_\theta}} \frac{\partial \theta}{\partial t'} +\xi_\theta. \label{eq_LangavineTheta2}
\end{equation}
where $t' = t / \sqrt{M_\theta}$. Therefore, the energy dissipated and then the line dividing the regions II and III of Fig. \ref{fig_map} scales with $\sqrt{M_\theta}$.

The value of $\gamma$ used in Figs. \ref{fig_Eff} and \ref{fig_EffPowervsTau} implies a terminal velocity of about $1/\tau = 3 \times 10^2 \mathrm{Hz}$ at $\delta \mu= 2 \times 10^{-4} V_0$. This was estimated from Eq. \ref{eq_Wtotal} with $W^{\mathrm{total}}=0$, assuming a constant terminal velocity, and $W_{\mathrm{(load)}}=0$. This velocity is the same than that of the nanomotor reported in Ref. \cite{FastNM}.

Considering the size of the system, $L=2000a$, the value of $\lambda$ used, $\lambda=25a$, ensures many periods of the potential within the system. Under this conditions a “gap” arises for Fermi energies $\varepsilon$ between $E_0 \pm \Delta$, where the energy of the center of the gap results in $E_0 = 2 V_0 - 2 V_0 \cos ( \pi a / \lambda )$. In finite systems and for energies within the band gap, the transmittance decays exponentially according to $T \approx e^{-2 L/\ell}$, where $\ell=\hbar v_F /\Delta$.\cite{AQM13} The Fermi velocity can be estimated for the tight-binding model at the center of the gap as $v_F=(2 V_0 a/\hbar) \sin(\pi a/\lambda)$.\cite{PM01}
Then, the decay length yields $\ell \approx 250 a$, for $\Delta=1 \times 10^{-3} V_0$.

The dwell time $\tau_D$ of electrons is not direct to evaluate as in the case of quantum dots, where this can be done from the width of the resonances. In the present case, electrons pass through the system by quantum tunneling and assessing tunneling times is a controversial and longstanding topic. However, just for the sake of comparing orders of magnitude we will use the definition of $\tau_D$ due to Smith,\cite{dwellTime,tunnelingT}
\begin{equation}
\tau_D = \frac{1}{v_F(k)} \int_{0}^{L} dx \left | \Psi_S(x,k) \right |^2
\end{equation}
where $\Psi(x,k)_S$ is the wave function inside the system.
We assume the wave function of the left lead $\Psi_L$ is $\Psi_L=e^{ikx}+r e^{-i k  x}$, where the reflection coefficient $r$ results in $r=-i e^{i \theta}$, according to the ideal model of the Thouless motor, for an energy at the center of the gap and assuming $L \rightarrow \infty$.\cite{AQM13} The wavelength $k$ is in this case $k=\pi/\lambda$.
For simplicity, let us also assume $\Psi_S \propto e^{-x/ \ell}$, where $\ell$ is the decay length. Then, the maximum value of the dwell time, which depends on the motor's coordinate $\theta$, is $\tau_D=2 \ell/v_F$. A similar result is obtained for the tunneling through a rectangular barrier in the limit of $L \rightarrow \infty$, $\tau_D= \frac{2 \ell}{v_F} \frac{1}{1+1/(\ell \ k)^2}$.\cite{tunnelingT}
With both formulas the dwell time gives $\tau_D \approx 3\times10^{-14} s$, which is completely negligible compared with the period of the motor rotating at maximum speed estimated above, $\tau \approx 3\times10^{-3}s$.

\bibliographystyle{unsrt}
\bibliography{./paper_TQM-postGalley}

\end{document}